\documentclass{amsart}
\usepackage{amsxtra}
\theoremstyle{plain}

\newcommand{\cleqn}{\setcounter{equation}{0}}
\newcommand{\clth}{\setcounter{theorem}{0}}
\newcommand {\sectionnew}[1]{\section{#1}\cleqn\clth}
\newcommand{\nn}{\hfill\nonumber}
\newtheorem{theorem}{Theorem}[section]
\newtheorem{lemma}[theorem]{Lemma}
\newtheorem{definition-theorem}[theorem]{Definition-Theorem}
\newtheorem{proposition}[theorem]{Proposition}
\newtheorem{corollary}[theorem]{Corollary}
\newtheorem{definition}[theorem]{Definition}
\newtheorem{example}[theorem]{Example}
\newtheorem{remark}[theorem]{Remark}
\newcommand \bth[1] { \begin{theorem}\label{t#1} }
\newcommand \ble[1] { \begin{lemma}\label{l#1} }

\newcommand \bpr[1] { \begin{proposition}\label{p#1} }
\newcommand \bco[1] { \begin{corollary}\label{c#1} }
\newcommand \bde[1] { \begin{definition}\label{d#1}\rm }
\newcommand \bex[1] { \begin{example}\label{e#1}\rm }
\newcommand \bre[1] { \begin{remark}\label{r#1}\rm }

\newcommand {\eth} { \end{theorem} }
\newcommand {\ele} { \end{lemma} }

\newcommand {\epr} { \end{proposition} }
\newcommand {\eco} { \end{corollary} }
\newcommand {\ede} { \end{definition} }
\newcommand {\eex} { \end{example} }
\newcommand {\ere} { \end{remark} }

\newcommand \thref[1]{Theorem \ref{t#1}}
\newcommand \leref[1]{Lemma \ref{l#1}}
\newcommand \prref[1]{Proposition \ref{p#1}}

\newcommand \deref[1]{Definition \ref{d#1}}

\newcommand \lb[1]{\label{#1}}
\def \d {{\partial}}   
\def \dd {\mathrm{d}}
\def \Cset {{\mathbb C}}
\def \Zset {{\mathbb Z}}

\def \Wr {{\mathcal{W}}_{rat}}
\def \Wp {{\mathcal{W}}_{poly}}
\def \B  {{\mathcal{B}}}
\def \C  {{\mathcal{C}}}
\def \D  {{\mathcal{D}}}
\def \SD  {{\mathcal{SD}}}
\def \S  {{\mathcal{S}}}

\def \Ga {\Gamma}

\def \Ps {\Psi}
\def \vp {\phi}

\def \mt  {\mapsto}
\def \ra  {\rightarrow}           

\def \sub {\subset}

\def \o  {\otimes}
            
\def \wt {\widetilde}

\def \id { {\mathrm{id}} }

\DeclareMathOperator \Span { {\mathrm{Span}} }  
\DeclareMathOperator \ord { {\mathrm{ord}} }

\begin{document}
\title[The prolate spheroidal phenomena and bispectrality]
{The prolate spheroidal phenomena and bispectrality}
\author[F. Alberto Gr\"unbaum]{F. Alberto Gr\"unbaum}
\address{
Department of Mathematics \\
University of California at Berkeley \\
Berkeley, CA 94720, U.S.A.}
\email{grunbaum@math.berkeley.edu}
\author[Milen Yakimov]{Milen Yakimov}
\address{
Department of Mathematics \\
Cornell University \\
Ithaca, NY 14853, U.S.A.}
\email{milen@math.cornell.edu}
\date{}
\begin{abstract}
Landau, Pollak, Slepian, and Tracy, Widom discovered that certain integral 
operators with so called Bessel and Airy kernels possess 
commuting differential operators and found important applications of this
phenomena in time-band limiting and random matrix theory.
In this paper we announce that very large classes of integral
operators derived from bispectral algebras of rank $1$ and $2$
(parametrized by lagrangian grassmannians of infinitely large size)  
have this property. The above examples come from special points in these
grassmannians.

\end{abstract}
\maketitle
\sectionnew{Introduction}\lb{Intro}
It was discovered by Landau, Pollak, Slepian \cite{SP, LP1, LP2, S}
and Tracy, Widom \cite{TW1, TW2}
that certain integral operators associated to the Airy and Bessel special 
functions posses commuting differential operators. They found important 
applications of this to time-band limiting, and to the study of 
asymptotics of Fredholm determinants, relevant to scaling limits of random 
matrix models. We call this phenomena the prolate spheroidal phenomena.

On the other hand,
the problem of bispectrality was posed \cite{DG} about 20 years ago by one 
of us (A. G.) and J. J. Duistermaat as a tool to understand this prolate 
spheroidal property of 
integral operators. The aim was to extend it to larger classes and
search for possible applications. Despite the dramatic
recent developments
in the areas of random matrices and bispectrality the two 
problems remained isolated except for several common examples, 
see \cite{G1, G2, G3}. In addition only few integral operators possessing 
the prolate spheroidal phenomena were found.

Here we announce that any selfadjoint bispectral algebra of ordinary 
differential operators (see \deref{Airy_Darboux} and \deref{selfadBess} 
and Section 2 for 
general definitions) of rank $1$ and $2$  
induces an integral operator possessing the prolate spheroidal 
property. The kernel of such an operator is of the form 
\[
K(x, z) = \int_{\Ga_2} \Psi(x, z) \Psi(y, z) dz
\]
where $\Psi(x,z)$ is the corresponding bispectral (eigen)function and  
$\Ga_2$ is a contour in the complex plane with $1$ or $2$ end points. It 
acts on the space $L^2(\Ga_1)$ again 
for a contour with the same property. The main results are stated in
\thref{FinalA} and \thref{FinalB}. 

The integral operators of Landau, Pollak, Slepian, and Tracy, Widom
correspond to the cases $\Psi(x,z) = \sqrt{xz} J_{\nu+1/2}(ixz)$ and
$\Psi(x, z) = A(x+z)$ which are known to be ``basic'' bispectral
functions, in the sense that the other rank $1$ and $2$ bispectral 
functions are obtained from them by certain type of Darboux 
transformations, see \cite{DG, W, BHYcmp, BHYphl, KR}. 
(Here $J_\nu(.)$ denote the Bessel functions of first kind and
$A(.)$ denotes the Airy function.) Although 
in these cases \cite{SP, LP1, LP2, S, TW1, TW2} the commuting differential
operator is of order 2, in general it is of arbitrarily large 
order.
 
In the rest of the introduction we describe our strategy for
proving \thref{FinalA} and \thref{FinalB} which relies 
on a very interesting property of the ``size'' of bispectral algebras
of rank $1$ and $2.$
 
Consider, more generally, a holomorphic function $\Psi(x,z)$ in some 
domain of $\Cset \times \Cset$ which is not an eigenfunction of any 
differential 
operator in $x$ or $z.$ Denote by $\B_\Ps$ the algebra of differential 
operators $R(x, \d_x)$ with rational coefficients for which there exists 
an operator $S(z, \d_z)$ with rational coefficients such that
\begin{equation}
\label{binvol1}
R(x, \d_x) \Ps(x, z) = S(z, \d_z) \Ps(x, z).
\end{equation}
The algebra of all differential operators $S(z, \d_z)$ obtained in this
way will be denoted by $\C_\Ps.$ The equality 
\[
b_\Psi (R(x, \d_x)) := S(z, \d_z)
\]
correctly defines an antiisomorphism from $\B_\Ps$ to $C_\Ps.$
Recall that such a function $\Psi(x,z)$ is called bispectral if both
algebras $\B_\Psi$ and $\C_\Psi$ contain rational 
functions. The subalgebra of $\B_\Psi$ and $\C_\Psi,$
consisting of differential operators in $x$ and $z$
for which $\Psi(x,z)$ is an eigenfunction, are commutative.
Algebras obtained in this way are called bispectral algebras.
Their rank (which is equal and is also called rank of the bispectral
function $\Psi(x,z))$ is the greatest common divisor of the orders 
of the operators of these algebras.

We derive our main result from the following remarkable property:

{\em{Consider the $\Zset_+ \times \Zset_+$ filtration of 
the algebra $\B_\Psi$ given by
\[
\B_\Psi^{l_1, l_2} 
= \{ R(x, \d_x) \in \B_\Psi \mid
\ord R(x, \d_x) \leq 2 l_1, \; \ord (b_\Psi R)(x, \d_x) \leq 2 l_2 \}.
\] 

If $\Psi(x,z)$ is a bispectral function of rank $r=1$ or $2$ then
the the size of the spaces $\B_\Psi^{l_1, l_2}$ in this filtration 
is large in the sense
\[
\dim \B_\Psi^{l_1, l_2} \geq \frac{2}{r}( 2 l_1 l_2 + l_1 + l_2) 
- conts
\]
where the constant is independent of $l_1$ and $l_2.$}}

For the basic bispectral functions $e^{xz}$ of rank $1,$ and
$A(x+z),$ 
$\sqrt{xz} J_{\nu+1/2}(ixz),$ $\nu \in \Cset \backslash \Zset$
of rank 2, the dimension of these spaces is exactly equal to the 
right hand side
with $const=-1.$ This is remarkable since it shows that 
for all bispectral functions $\Psi(x,z)$ of rank 1 and $2$ the 
spaces of the above natural filtration of the algebra $\B_\Psi$
are approximately of the same dimension as the basic (Bessel and Airy)
ones.

It is very interesting to understand the relation between our results
and the approach of Adler, Shiota, and van Moerbeke \cite{AvM, ASvM} to
the Tracy--Widom system of differential equations via Virasoro 
constraints. 
One possible way would be to incorporate the representation theoretic 
meaning 
of the Bessel tau functions from \cite{BHYd} in terms of representations 
of the $W_{1+\infty}$ algebra, see \cite{FKRW}.

Another important problem is to understand the relation of this work 
to the isomonodromic deformations approach to random matrices from the 
works of Palmer \cite{P}, Harnad--Tracy--Widom \cite{HTW, TW3}, 
Its--Harnad \cite{HI}, and Borodin--Deift \cite{BD}.

The proofs of the results announced here will appear in a forthcoming
publication.

Acknowledgments: M. Y. would like to thank the organizers of the
Workshop on Superintegrability, September 16--21, CRM, Montreal,
and especially John Harnad and
Pavel Winternitz for the opportunity to participate at the conference 
and for their interest in this work.
\sectionnew{Bispectrality and commutativity}
Denote by $\Wr$ the algebra of differential operators in one variable with 
rational coefficients. Denote by $a(.)$ the formal adjoint of 
operators in $\Wr:$
\begin{equation}
\label{adjoint}
a \left(\sum_{k=0}^n b_k(x) \d_x^n \right) =
\sum_{k=0}^n (-\d_x)^n b_k(x).
\end{equation}
For any $\xi \in \Cset$ consider the map from the subalgebra 
of $\Wr$ consisting of operators regular at $\xi$ to the 
set of bidifferential operators at $\xi$
\begin{equation}
\label{boundary_oper}
\vp_\xi \left(\sum_{k=0}^n b_k(x) \d_x^n \right) =
\sum_{k=0}^n \sum_{i=0}^{k-1} (-1)^i \d_x^{k-i-1} \o \d_x^i b_k(x)
\Big|_{x=\xi}.
\end{equation}

For an oriented contour $\Ga$ in $\Cset$ we will denote its endpoints 
by $e(\Ga).$ For any $\xi \in e(\Ga)$ set $\pi(\xi)=1$ or $0$ depending on 
whether $\xi$ is a left or right endpoint of $\Ga.$ 

Assume that 
\[
D(x, \d_x) = \sum_{k=0}^n b_k(x) \d_x^n  \in \Wr
\]
is a differential operator which is regular along $\Ga.$
Let $f(x)$ and $g(x)$ be smooth functions on $\Ga$ that decrease rapidly 
when $x \ra \infty$ (i.e. $\lim_{x \to \infty} p(x) f^{(k)}(x)=0$ for any 
polynomial $p(x)$ and any integer $k,$ similarly for $g(x)).$

By standard integration by parts one gets that
\begin{align}
\label{byparts}
\int_\Ga \left( D(x, \d_x) f(x) \right) g(x) \dd x  =
&\sum_{\xi \in e(\Ga)} (-1)^{\pi(\xi)} \vp_\xi(D(x, \d_x))(f(x) \o g(x)) 
\\
&+ \int_\Ga f(x) \left( aD(x, \d_x) g(x) \right) \dd x.
\nn
\end{align}

Let $\Psi(x,z)$ be a holomorphic function on some domain in 
$\Cset \times \Cset$ which is not an eigenfunction of a 
differential operator in $x$ or $z,$ as in the introduction. 
Recall the definition\eqref{binvol1} of the algebras of differential 
operators 
$\B_\Psi$ and $\C_\Psi.$
Assume that $\wt{\B}_\Ps$ and $\wt{\C}_\Ps$ are two subalgebras of 
$\B_\Ps$ and $\C_\Ps,$ respectively, that are stable under the 
formal adjoint map \eqref{adjoint} and such that 
\[
b_\Psi (\wt{\B}_\Ps) = \wt{\C}_\Ps.
\]

\bpr{commut} Let $\Ga_1$ and $\Ga_2$ be two contours in 
$\Cset$ such that $\Ga_1 \times \Ga_2$ is in the domain of 
$\Ps(x, z)$ and $\Ps(x, z)$ decreases rapidly when $x$ or $z$ go to 
$\infty$ along 
$\Ga_1$ and $\Ga_2.$ If $D(x, \d_x) \in \wt{\B}_\Ps$ is such that
\begin{equation}
\label{ab}
a b_\Psi (D(x, \d_x)) = b_\Psi a(D(x, \d_x))
\end{equation}
and
\begin{equation}
\label{boundary}
\vp_\xi(D(x, \d_x)) = 0, \, \forall \xi \in e(\Ga_1); \quad
\vp_\xi((bD)(z, \d_z)) = 0, \, \forall \xi \in e(\Ga_2)
\end{equation}
then the integral operator with kernel
\begin{equation}
\label{ker}
K(x,y) = \int_{\Ga_2} \Ps(x,z) \Ps(y,z) \dd z
\end{equation}
on $L^2(\Ga_1)$ commutes with the differential operator
$D(x, \d_x)$ with domain all smooth functions on $\Ga_1$ 
that decrease rapidly as $x \ra \infty.$
\epr

Let us call a differential operator 
$D(x, \d_x) \in \Wr$ 
{\em{formally symmetric}} if
\[
(a D)(x, \d_x) = D(x, \d_x)
\]
and {\em{formally skewsymmetric}} if 
\[
(a D)(x, \d_x) = - D(x, \d_x).
\]

\ble{symm} (i) A differential operator 
$D(x, \d_x) \in \Wr$ is formally symmetric if and only if it has the
form
\begin{equation}
\label{symm}
D(x, \d_x) = \sum_{i=0}^n \d_x^n c_i(x) \d_x^n
\end{equation}
for some integer $n$ and some rational functions $c_i(x).$

(ii) The operator $D(x, \d_x)$ given by \eqref{symm} satisfies
\[
\vp_\xi(D(x, \d_x))=0
\]
for some fixed $\xi \in \Cset$ if and only if
\[
(\d_x^i c_k)(\xi)= 0 \,
\mbox{for} \, k=1, \ldots, n, \, 
i=0, \ldots, k-1.
\]
\end{lemma}

For a given function $\Ps(x,z)$ as before let 
$\B_{\Ps,sym}$ be the subalgebra of $\B_\Ps$ consisting of 
differential operators $R(x, \d_x)$ for which both
$R(x, \d_x)$ and $(b_\Psi R)(z, \d_z)$ are formally symmetric.
Set also $\C_{\Ps,sym} := b_\Psi (\B_{\Ps,sym}).$
Define the vector spaces
\begin{align}
\B_{\Ps,sym}^{l_1, l_2} &= \{ R(x, \d_x) \in \B_{\Ps,sym} \mid
\ord R(x, \d_x) \leq 2 l_1 \, \mbox{and} \, 
\ord (b_\Psi R)(z, \d_z) \leq 2 l_2
\},
\label{Bsym} \\
\C_{\Ps,sym}^{l_1, l_2} &= \{ S(z, \d_z) \in \C_{\Ps,sym} \mid
\ord (b_\Psi^{-1}S)(x, \d_x) \leq 2 l_1 \, 
\mbox{and} \ord S(z, \d_z) \leq 2 l_2
\}.
\label{Csym}
\end{align}
Clearly $\C_{\Ps, sym}^{l_1, l_2}= b_\Psi (\B_{\Ps, sym}^{l_1, l_2}).$

{} From \prref{commut} and \leref{symm} we obtain:

\bth{com_cond} Assume the conditions from \prref{commut} for the the 
function $\Ps(x, z)$ and the contours $\Ga_1$ and $\Ga_2.$ If either of 
the following two conditions is satisfied  
then the integral operator with kernel \eqref{ker} possesses a 
formally commuting symmetric differential operator of order less than or 
equal to 
$2l_1$ and domain -- the space of rapidly decreasing smooth functions on 
$\Ga_1.$

Condition (i): 
\[
\B_{\Ps,sym}^{l_1, l_2} >
l_1(l_1+1) e(\Ga_1)/2 +  l_2(l_2+1) e(\Ga_2)/2,
\]

Condition (ii): $-e(\Ga_1)=e(\Ga_1),$ $-e(\Ga_2)=e(\Ga_2),$ all operators
in $\B_{\Ps,sym}$ are invariant under the transformation $x \mt -x,$
and
\[
\B_{\Ps,sym}^{l_1, l_2} >
l_1(l_1+1) e(\Ga_1)/4 +  l_2(l_2+1) e(\Ga_2)/4   
\]
\eth

\sectionnew{Integral operators associated to selfadjoint Darboux 
transformations of Airy functions}
\subsection{The Airy bispectral function}
Denote by $A(x)$ the Airy function and set
\begin{equation}
\label{Airy}
\Psi_A(x, z) = A(x+z).
\end{equation}
Recall that $A(x)$ decreases rapidly when $x \ra \infty$ in the sector
$ - \pi/3 < \arg x < \pi/3.$

If $L_A(x, \d_x)$ denotes the Airy differential operator
\[
L_A(x, \d_x) = \d_x^2 -x
\]
then $\Psi_A(x, z)$ satisfies 
\begin{align}
L_A(x, \d_x) \Psi_A(x, z) &= z \Psi_A(x, z),
\label{AA1} 
\\
\d_x \Psi_A(x, z) &= \d_z \Psi_A(x, z),
\label{AA2} 
\\
x \Psi_A(x, z) &= L_A(z, \d_z) \Psi_A(x, z). 
\label{AA3}
\end{align}
For shortness denote the algebras $\B_{\Psi_A}$
and $\C_{\Psi_A}$ of differential operators with rational coefficients
associated to the Airy function $\Psi_A(x,z),$
recall \eqref{binvol1}, by $\B_A$ and $\C_A.$ It is straightforward to 
deduce:

\ble{BCAiry} The algebras $\B_A$ and $\C_A$ coincide with the Weyl algebra
$\Wp$ of differential operators in one variable with polynomial 
coefficients. Moreover the antiisomorphism $b_A$ associated to the Airy 
function $\Psi_A(x, z),$ recall \eqref{Airy}, is uniquely defined from the 
relations
\[
b_A(x)= (L_A(z, \d_z)), \quad
b_A(\d_x) = \d_z, \quad
b_A(L_A(x, \d_x)) = z.
\]
\ele

\subsection{Selfadjoint Darboux transformations from the Airy function}
Note that 
\[
\Cset[x] = \B_A \cap \Cset(x)
\]
and 
\[
\Cset[L_A(x, \d_x)] = b^{-1}_A ( \C_A \cap \Cset(z) ).
\]

The set of rational Darboux transformations $\D_A$ from the Airy 
function was defined in 
\cite{BHYcmp} as the set of functions $\Psi(x, z)$ for which 
there exist differential operators 
\begin{equation}
P(x, \d_x), Q(x, \d_x) \in
(\B_A)_{(\Cset[x] \backslash 0)} = \Wr
\label{localize}
\end{equation}
such that
\begin{align}
f(L_A(x, \d_x) ) &= Q(x, \d_x) P(x, \d_x),
\label{DT1} \\
\Psi(x, z) & = \frac{1}{p(z)}P(x, \d_x) \Psi_A(x, z),
\label{DT2}
\end{align}
for some polynomials $f(t)$ and $p(z).$ (The polynomial $p(z)$ is included
for normalization purposes only.) The quotient ring of $\B_A$ by 
$\Cset[x] \backslash \{0 \}$ 
in \eqref{localize} is well defined since 
$\Cset[x] \backslash \{0 \}$ satisfies the Ore condition,
see \cite{MR}.

It was also shown in \cite{BHYcmp, KR} and more conceptually
proved in \cite{BHYphl} that:

\bth{Airybisp} All rational Darboux transformations from the 
Airy function $\Psi(x, z)$ are bispectral functions of rank $2.$
\eth

\bde{Airy_Darboux} Define the set $\SD_A$ of selfadjoint Darboux 
transformations from the Airy function $\Psi(x, z)$ to consist of those 
functions $\Psi(x, z)$ for which there exists a 
differential operator $P(x, \d_x) \in \Wr$ such that
\begin{align}
g(L_A(x, \d_x))^2 &= (aP)(x, \d_x) P(x, \d_x),
\label{saDT1} \\
\Psi(x, z) &= \frac{1}{g(z)} P(x, \d_x) \Psi_A(x, z),
\label{saDT2}
\end{align}
for some polynomial $g(t).$ 
\ede

In fact $\SD_A$ consists exactly of those $\Psi(x, z) \in \D_A$ for which 
$Q(x, \d_x) = (aP) (x, \d_x)$ in \eqref{DT1}--\eqref{DT2}
with an appropriate normalization of the polynomial $p(z).$
One can show that as a consequence $f(t)$ is the square of some 
polynomial $g(t),$ compare to \eqref{saDT1}--\eqref{saDT2}.

\subsection{Size of the algebra $\B_A$ relative to the antiisomorphism 
$b_A$}
Consider the $\Zset_+ \times \Zset_+$ filtration of the algebra
$\B_A$ associated to the Airy function $\Psi_A(x, z)$ 
defined analogously to \eqref{Bsym} by
\begin{equation}
\B_A^{l_1, l_2} = \{ R(x, \d_x) \in \B_A \mid 
\ord R(x, \d_x) \leq 2 l_1, \ord (b_A R)(z, \d_z) \leq 2 l_2
\}.
\label{Airy12}
\end{equation}

\ble{B12} The vector space $\B_A^{l_1, l_2}$ has a basis 
that consists of the differential operators
\[
\{ x^m (L_A(x, \d_x))^n \mid n \leq l_1, m \leq l_2 \} 
\cup
\{ x^m \d_x (L_A(x, \d_x))^n \mid n < l_1, m < l_2 \}.
\]
\ele

Note that the formal adjoint antiinvolution $a$ of $\Wr$ preserves
the spaces $\B_A^{l_1, l_2}.$ Since $a^2 = \id$ the space 
$\B_A^{l_1, l_2}$ is the direct sum of the eigenspaces of $a$ with 
eigenvalues $\pm1.$ Denote the eigenvalue $1$ subspace of 
$\B_A^{l_1, l_2}$ by $\B_{A, sym}^{l_1, l_2}.$ For the Airy function 
$\Psi_A(x, z)$ one has $a b_A =b_A a$ and thus:
\begin{align}
\B_{A, sym}^{l_1, l_2}= \{ R(x, \d_x) \in \B_A \mid
&\ord R(x, \d_x) \leq 2 l_1, \, \ord (b_A R)(z, \d_z) \leq 2 l_2,
\nn \\
&aR(x, \d_x) =R(x, \d_x), \, a(b_A R(x, \d_x)) = b_A R(x, \d_x) \}.
\nn
\end{align}

\ble{B12a} The set of operators
\begin{equation}
\{ x^m (L_A(x, \d_x))^n + (L_A(x, \d_x))^n x^m 
\mid n \leq l_1, m \leq l_2 \} 
\label{basisA+}
\end{equation}
is a basis for the space $\B_{A, sym}^{l_1, l_2}.$ 
In particular 
\[
\dim \B_{A, sym}^{l_1, l_2} = (l_1+1)(l_2+1).
\]
\ele

\subsection{Size of the algebra $\B_{\Psi, sym}$  
for a selfadjoint Darboux transformation from the Airy function,
relative to the involution $b_\Psi$}
Fix an arbitrary selfadjoint Darboux transformation from the Airy 
function  
$\Psi \in \SD_A$ given by \eqref{saDT1}--\eqref{saDT2} for some 
$P(x, \d_x) \in \Wr$ and $g(t) \in \Cset[t].$ Let
\begin{equation}
\label{Va}
P(x, \d_x) = \frac{1}{v(x)} R(x, \d_x)
\end{equation}
for some $R(x, \d_x) \in \Wp = \B_A$ and $v(x) \in \Cset[x].$
Set
\begin{equation}
\label{orders}
\ord R(x, \d_x) = \rho_1 \quad \mbox{and} \quad
\ord (b_A R)(x, \d_x) = \rho_2.
\end{equation}
Denote
\begin{align}
\S_{\Psi, 1} &= \Span \{ \frac{1}{v(x)} R(x, \d_x) M(x, \d_x) (a R)(x, \d_x)
\frac{1}{v(x)} \mid M(x, \d_x) \in \B_{A, sym}^{l_1 - \rho_1, l_2} \},
\nn \\
\S_{\Psi, 2} &= \Span \{ v(x) M(x, \d_x) v(x)  
\mid M(x, \d_x) \in \B_{A, sym}^{l_1, l_2 - \rho_2} \}.
\nn
\end{align}

\bpr{Atr1} In the above setting:

(i) The spaces of differential operators $\S_{\Psi, 1}$ and 
$\S_{\Psi, 2}$ are subspaces of $\B_{\Psi, sym}^{l_1, l_2}.$

(ii) The dimension of the intersection $\S_{\Psi, 1} \cap \S_{\Psi, 2}$ is 
less than or equal to
\[
(l_1 -\rho_1 + 1)(l_2 -\rho_2 +1).
\]
\epr

Theorem 4.2 of \cite{BHYphl} shows that \eqref{saDT1}--\eqref{saDT2}
is equivalent to  
\begin{align}
v(L_A(z, \d_z))^2 &= (a b_A R)(z, \d_z) \frac{1}{g(z)^2} (b_A R)(z, \d_z),
\nn
\\
\Psi(x, z) &= \frac{1}{v(x) g(z)} (b_A R)(z, \d_z) \Psi_A (x, z).
\nn
\end{align}
(The hard step is to prove the first equality.) One can show that
\[
b_\Psi \left(
\frac{1}{v(x)} R(x, \d_x) M(x, \d_x) (aR)(x, \d_x) 
\frac{1}{v(x)} 
\right)
= g(z) (b_A M)(z, \d_z) g(z)
\]
for all operators $M(x, \d_x) \in \B_A.$ The fact that 
$\S_{\Psi,2}$ is a subspace of $\B_{\Psi, sym}^{l_1, l_2}$ 
follows from this and the fact that 
$\S_{\Psi, 1} \sub \B_{\Psi, sym}^{l_1, l_2}$
by exchanging the roles of $x$ and $z.$

\bth{Atr} For any selfadjoint Darboux transformation from the 
Airy function $\Psi(x, z) \in \SD_A$ the dimension of the space
of differential operators $\S_{\Psi, 1} + \S_{\Psi, 2}$ is greater than 
or equal to $(l_1 +1)(l_2 +1) - \rho_1 \rho_2.$ In particular  
\begin{equation}
\label{Amain}
\dim \B_{\Psi, sym}^{l_1, l_2} \geq
(l_1 +1)(l_2 +1) - \rho_1 \rho_2.
\end{equation}
\eth

\thref{Atr} and \thref{com_cond} imply our final result for integral
operators derived from Darboux transformations from the Airy function.

\bth{FinalA} Let $\Psi(x, z) \in \SD_A$ be a selfadjoint Darboux 
transformation from the Airy function, given by \eqref{saDT1}, 
\eqref{saDT2}, \eqref{Va}. Let $\Ga_1,$ $\Ga_2$ be two connected contours 
in $\Cset$ that do not contain the roots of the polynomials
$v(t)$ and $g(t)$ respectively and that begin at some finite points and 
go to infinity in the sector $- \pi/3 < \arg x < \pi/3.$ Then the integral 
operator
\[
K(x,y) = \int_{\Ga_2} \Ps(x,z) \Ps(y,z) \dd z
\]
on $L^2(\Ga_1)$ commutes with a formally symmetric differential operator 
with rational coefficients of order less than or equal to 
$2 \rho_1 \rho_2$ 
and domain all smooth functions on $\Ga_1$ that decrease rapidly as 
$x \ra \infty.$
\eth
\sectionnew{Integral operators associated to selfadjoint Darboux 
transformations of Bessel functions}
\subsection{The Bessel bispectral function}
Denote by  $J_\nu$ the standard Bessel functions of first kind. 
By abuse of notation the functions 
\begin{equation}
\label{Besself}
\Psi_\nu(x,z) = (xz)^{1/2} J_{\nu+1/2}(ixz)
\end{equation}
will be also called Bessel functions. Consider the Euler operator
\[
D_x = x \d_x 
\]
and the operators
\begin{equation}
\label{Besseloper}
L_\nu(x, \d_x) = \d^2_x - \frac{\nu(\nu+1)}{x^2} =
\frac{1}{x^2} (D_x + \nu)(D_x - \nu - 1)
\end{equation}
to be called Bessel operators. The Bessel functions satisfy the equations
\begin{align}
\label{B1}
L_\nu(x, \d_x) \Psi_\nu(x, z) &= z^2 \Psi_\nu(x, z),
\\
\label{B2}
D_x \Psi_\nu(x, z) &= D_z \Psi_\nu(x, z),
\\
\label{B3}
x^2 \Psi_\nu(x, z) &= L_\nu(z, \d_z) \Psi_\nu(x, z).
\end{align}
For shortness the algebras $\B_{\Psi_\nu}$ and $\C_{\Psi_\nu}$ associated 
to the Bessel function $\Psi_\nu(x, z),$ recall \eqref{binvol1},
will be denoted by $\B_\nu$ and $\C_\nu.$ 

The Bessel functions corresponding to $\nu_1, \nu_2 \in \Cset$ that differ 
by an integer can be obtained by a Darboux transformation from each other:
\begin{align}
\Psi_{\nu+1}(x,z) = \frac{1}{xz} (D_x - \nu - 1 ) \Psi_\nu(x, z),
\nn \\
\Psi_\nu(x,z) = \frac{1}{xz} (D_x + \nu +1) \Psi_{\nu+1}(x, z),
\nn 
\end{align}
which corresponds to the factorizations
\[
L_\nu= x^{-1} (D_x + \nu + 1 ) x^{-1} (D_x - \nu -1 ), \quad
L_{\nu+1}= x (D_x - \nu -1 ) x^{-1} (D_x + \nu + 1).
\]

According to \eqref{B1}--\eqref{B3} the algebras of differential operators
$\B_\nu$ and $\C_\nu$ contain the operators $L_\nu(x, \d_x),$ $D_x,$ and 
$x^2.$ Denote their subalgebras generated by those operators by 
$\wt{\B}_\nu$ and $\wt{\C}_\nu,$ respectively. Clearly 
\eqref{B1}--\eqref{B3} imply
\[
b_\nu(\wt{\B}_\nu)=\wt{\C}_\nu.
\]
Similarly to Proposition 2.4 in \cite{BHYcmp} one shows:

\ble{BesselBC} (i) For $\nu \in \Cset \backslash \Zset$ the algebras 
$\B_\nu$ and $\C_\nu$ are generated by the operators $L_\nu(x, \d_x),$ 
$D_x,$ and $x^2,$ i.e.  
\[
\B_\nu = \wt{\B}_\nu, \quad 
\C_\nu = \wt{\C}_\nu.
\]

(ii) For $\nu \in \Zset$ the subalgebras $\wt{\B}_\nu$ and
$\wt{\C}_\nu$ of $\B_\nu$ and $\C_\nu$ consist of exactly those 
differential operators in $\B_\nu$ and $\C_\nu$ that are invariant under 
the transformation $x \mt -x.$ 
\ele

Note that
\begin{equation}
a b_\nu (P(x, \d_x)) = b_\nu a (P(x, \d_x)), \; 
\forall P(x, \d_x)  \in \wt{\B}_\nu.
\label{abBess}
\end{equation}

\subsection{Selfadjoint Darboux transformations from the Bessel functions}
Similarly to the Airy case we have
\begin{align}
\Cset[x^2] &= \wt{\B}_\nu \cap \Cset(x),
\nn \\
\Cset[L_\nu(x, \d_x)] &= b^{-1}_\nu (\wt{\C}_\nu \cap \Cset(z)).
\nn 
\end{align}

The set of rational Darboux transformations $\D_\nu$ from the 
Bessel function $\Psi_\nu(x, z)$ is defined to be the set of functions 
$\Psi(x, z)$ for which there exist differential operators 
\begin{equation}
P(x, \d_x), Q(x, \d_x) \in 
(\B_\nu)_{(\Cset[x^2] \backslash \{0\})} 
\label{pqBes}
\end{equation}
such that
\begin{align}
f(L_\nu(x, \d_x)) &= Q(x, \d_x) P(x, \d_x),
\label{ratB1} \\
\Psi(x, z) &= \frac{1}{p(z)} P(x, \d_x) \Psi_\nu(x, z),
\label{ratB2}
\end{align}
for some polynomials $f(t)$ and $p(z).$ 
Again the subset $\Cset[x^2] \backslash \{0\}$ of $\B_\nu$
satisfies the Ore condition and the quotient ring in \eqref{pqBes} 
makes sense.

The following theorem was proved in \cite{W} for $\nu =1$ and in  
\cite{BHYcmp, BHYphl} in the general case.
\bth{bispBess} All rational Darboux transformations from the Bessel 
functions are bispectral of rank $2$ if $\nu \in \Cset \backslash \Zset$
and of rank 1 if $\nu \in \Zset.$
\eth

\bde{selfadBess} Define the set of selfadjoint (``even, selfadjoint'' 
in the case $\nu \in \Zset)$ Darboux transformations $\SD_\nu$ from the 
Bessel functions 
$\Psi_\nu(x, z)$ to consist of all functions $\Psi(x,z)$ for 
which there exists a differential operator  
\[
P(x, \d_x) \in (\Cset[x^2] \backslash \{0\})^{-1} \wt{\B}_\nu
\]
such that
\begin{align}
f(L_\nu(x, \d_x)) &= (-1)^m (aP)(x, \d_x) P(x, \d_x)
\label{saDT1B} 
\\
\Psi(x, z) &= \frac{1}{z^m g(z^2)} P(x, \d_x) \Psi_\nu(x, z),
\label{saDT2B} 
\end{align}
for a polynomial $f(t)$ of the form
\begin{equation}
f(t) = t^{2m} g(t^2)^2, \; g(0) \neq 0; g(t) \in \Cset[t].
\label{formf}
\end{equation}
\ede
{\em{Here the term ``even'' reflects the fact that for $\nu \in \Zset$
the algebras $\B_\nu$ and $\C_\nu$ are bigger than 
$\wt{\B}_\nu$ and $\wt{\C}_\nu.$}} 
The reason for this terminology is explained below.

As in the Airy case the set $\SD_\nu$ consists of those rational 
Darboux transformations $\Psi(x, z)$ from the Bessel function
$\Psi_\nu(x,z)$ for which in the notation \eqref{ratB1}--\eqref{ratB2}
\begin{equation}
Q(x, \d_x) = (-1)^{\ord P}(a P)(x, \d_x)
\label{QP}
\end{equation}
with the additional property in the case $\nu \in \Zset$
\begin{equation}
P(-x, -\d_x) = P(x, \d_x).
\label{even}
\end{equation}
As a consequence it is obtained that the polynomial $f(t)$ in 
\eqref{ratB1} has the form \eqref{formf} and an appropriate normalization 
of $p(z)$ is made.

Note that \leref{BesselBC} implies that in the rank $2$ case
$\nu \in \Cset \backslash \Zset$ the condition \eqref{even} is 
a consequence of \eqref{pqBes}. In the rank 1 case $\nu \in \Zset$ the 
term ``even'' 
reflects this extra condition. It is needed since in the case of 
Darboux transformations from the Bessel function the prolate spheroidal 
property will be deduced from the second condition in \thref{com_cond}.
\subsection{Size of the algebras $\wt{B}_\nu$ relative to the 
antiisomorphisms $b_\nu$}
Consider the $\Zset_+ \times \Zset_+$ filtration of the algebras
$\wt{B}_\nu$ 
\begin{equation}
\label{Bessel12}
\wt{\B}_\nu^{l_1, l_2} = \{ R(x, \d_x) \in \wt{\B}_\nu \mid 
\ord R(x, \d_x) \leq 2 l_1, \ord (b_\nu R)(z, \d_z) \leq 2 l_2
\}
\end{equation}

The formal adjoint involution $a$ of $\Wr$ preserves 
the spaces $\B_\nu^{l_1, l_2}.$ Similarly to \leref{B12} one shows:
\ble{BBes12} The vector space $\B_\nu^{l_1, l_2}$ has a basis 
that consists of the differential operators
\[
\{ x^{2m} (L_\nu(x, \d_x))^n \mid n \leq l_1, m \leq l_2 \} 
\cup
\{ x^{2m} D_x (L_\nu(x, \d_x))^n \mid n < l_1, m < l_2 \}.
\]
\ele

Since $a^2 = \id$ the space $\wt{\B}_\nu^{l_1, l_2}$ is the direct sum of 
the eigenspaces of $a$ with eigenvalues $\pm1.$ 
The eigenvalue $1$ subspace of $\wt{\B}_\nu^{l_1, l_2}$ will be denoted by 
$\wt{\B}_{\nu, sym}^{l_1, l_2}.$ The commutativity \eqref{abBess}
of $a$ and $b_\nu$ on $\wt{\B_\nu}^{l_1, l_2}$ implies:
\begin{align}
\wt{\B}_{\nu, sym}^{l_1, l_2}= \{ R(x, \d_x) \in \wt{\B}_\nu \mid
&\ord R(x, \d_x) \leq 2 l_1, \, \ord (b_\nu R)(z, \d_z) \leq 2 l_2,
\nn \\
&aR(x, \d_x) =R(x, \d_x), \, a(b_\nu R(x, \d_x) ) = b_\nu R(x, \d_x) \}.
\nn
\end{align}

\ble{B12aBess} The set of operators
\begin{equation}
\{ x^{2m} (L_\nu(x, \d_x))^n + (L_\nu(x, \d_x))^n x^{2m} 
\mid n \leq l_1, m \leq l_2 \} 
\label{basisA+Bess}
\end{equation}
is a basis for the space $\wt{\B}_{\nu, sym}^{l_1, l_2}.$ 
In particular 
\[
\dim \wt{\B}_{\nu, sym}^{l_1, l_2} = (l_1+1)(l_2+1).
\]
\ele

\subsection{Size of the algebra $\B_{\Psi, sym}$  
for an (even) selfadjoint Darboux transformation from a Bessel function}
Fix an arbitrary selfadjoint (and in addition even in the rank $1$ case
$\nu \in \Zset)$ Darboux transformation from a Bessel function
$\Psi_\nu(x,z),$ $\Psi \in \SD_\nu$ given by 
\eqref{saDT1B}--\eqref{saDT2B} for some 
$P(x, \d_x) \in (\Cset[x^2] \backslash \{0\})^{-1} \wt{\B}_\nu$ and 
$f(t)= t^{2m} g(t^2)^2,$ $g(t) \in \Cset[t].$ Let
\begin{equation}
\label{vB}
P(x, \d_x) = \frac{1}{v(x^2)} R(x, \d_x)
\end{equation}
for some $R(x, \d_x) \in \wt{\B}_\nu$ and $v(x) \in \Cset[x].$
Set
\begin{equation}
\label{ordersBes}
\ord R(x, \d_x) = \rho_1 \quad \mbox{and} \quad
\ord (b_\nu R)(x, \d_x) = \rho_2.
\end{equation}
Denote
\begin{align}
\S_{\Psi, 1} &= \Span \{ \frac{1}{v(x^2)} R(x, \d_x) M(x, \d_x) 
(a R)(x, \d_x) \frac{1}{v(x^2)} \mid M(x, \d_x) \in 
\wt{\B}_{\nu, sym}^{l_1 - \rho_1, l_2} \},
\nn \\
\S_{\Psi, 2} &= \Span \{ v(x^2) M(x, \d_x) v(x^2)  
\mid M(x, \d_x) \in \wt{\B}_{\nu, sym}^{l_1, l_2 - \rho_2} \}.
\nn
\end{align}

\bpr{Atr1B} (i) The spaces of differential operators $\S_{\Psi, 1}$ and 
$\S_{\Psi, 2}$ are subspaces of $\B_{\Psi, sym}^{l_1, l_2}$ and are
invariant under the transformation $x \mt -x.$

(ii) The dimension of the intersection $\S_{\Psi, 1} \cap \S_{\Psi, 2}$ is 
less than or equal to
\[
(l_1 -\rho_1 + 1)(l_2 -\rho_2 +1).
\]
\epr

Finally we obtain the following Theorem.

\bth{Btr} For any selfadjoint (and even, selfadjoint in the case
$\nu \in \Zset)$ Darboux transformation from a 
Bessel function $\Psi_\nu(x, z),$ $\Psi(x, z) \in \SD_\nu$ 
the space $\S_{\Psi, 1} + \S_{\Psi, 2}$ consists 
of differential operators invariant under the transformation $x \mt -x$
\[
\dim (\S_{\Psi, 1} + \S_{\Psi, 2}) \geq 
(l_1 +1)(l_2 +1) - \rho_1 \rho_2.
\]
In particular the dimension of the subspace of $\B_{\Psi, sym}^{l_1, l_2}$
of differential operators invariant under $x \mt -x$ is 
greater than or equal to $(l_1 +1)(l_2 +1) - \rho_1 \rho_2.$
\eth

\thref{Btr} and \thref{com_cond} imply our final result for Darboux 
transformations from the Bessel functions:

\bth{FinalB} Let $\Psi(x, z) \in \SD_\nu$ be a selfadjoint 
(and in addition even if $\nu \in \Zset)$ Darboux 
transformation from the Bessel function $\Psi_\nu(x,z),$
given by \eqref{saDT1B}, \eqref{saDT2B}, \eqref{formf},
\eqref{vB}. 
Let $\Ga_1,$ $\Ga_2$ be two connected finite contours 
that do not contain the roots of the polynomials $v(t)$ and $g(t),$ 
respectively and such that $e(\Ga_i)=-e(\Ga_i).$  

Then the integral operator with kernel
\[
K(x,y) = \int_{\Ga_2} \Ps(x,z) \Ps(y,z) \dd z
\]
on $L^2(\Ga_1)$ commutes with a formally symmetric differential operator 
with rational coefficients of order less than or equal to 
$2 \rho_1 \rho_2$ and domain all smooth functions on $\Ga_1.$ 
\eth

\bre{lin} Let $\Phi_1(x,z)$ and $\Phi_2(x,z)$ be two functions
satisfying \eqref{AA1}--\eqref{AA3} or 
\eqref{B1}--\eqref{B2}. E.g. in the rank 
$1$ case (section 4, $\nu=0$) one can choose 
\[
\Phi_1(x,z) = e^{xz}, \quad \Phi_2(x,z) = e^{-xz}
\]
and in the rank 2, Bessel case (section 4)
\[
\Phi_1(x,z) = \Psi_\nu(x,z), \quad \Phi_2(x,z) = \Psi_\nu(x,-z).
\]
Fix a differential 
operator $P(x, \d_x)$ and a polynomial $g(z)$ that define an
(even) selfadjoint Darboux transformation from the Airy or Bessel
functions as in \deref{Airy_Darboux} and \deref{selfadBess}. Define
\[
\Psi_i(x,z) = \frac{1}{g(z)} P(x, \d_x), \;
\mbox{and} \;
\Psi_i(x,z) = \frac{1}{z^mg(z^2)} P(x, \d_x), \; 
i=1,2
\]
in the Airy and Bessel cases, respectively. Then \thref{FinalA}
and \thref{FinalB} hold for the integral operator with kernel
\[
K(x,y) = \int_{\Ga_2} \Psi_1(x,z) \Psi_2(y,z) \dd z
\]
and exactly the same commuting differential operator that appears in 
those Theorems.
\ere

\end{document}